\documentclass{article}

\usepackage{epsfig} 
\newlength{\abstractwidth} 
\renewcommand{\thefootnote}{\fnsymbol{footnote}} 
\renewcommand{\thanks}[1]{\footnote{#1}} 
\newcommand{\starttext}{ 
\setcounter{footnote}{0} 
\renewcommand{\thefootnote}{\arabic{footnote}}} 
\renewcommand{\theequation}{\thesection.\arabic{equation}} 
\newcommand{\be}{\begin{equation}} 
\newcommand{\bea}{\begin{eqnarray}} 
\newcommand{\eea}{\end{eqnarray}} 
\newcommand{\ee}{\end{equation}} 
 
\newcommand{\Oplus}{$\mathbf{O}\llap+ \: $}

\def\12{{1 \over 2}}

\newcommand{\tab}{\hspace{5mm}}

\begin{document} 
\renewcommand{\theequation}{\thesection.\arabic{equation}} 
\begin{titlepage} 
\bigskip
\centerline{\Large \bf {Group Structure of an Extended Lorentz Group}}
\bigskip
\begin{center} 
{\large James Lindesay\footnote{
Permanent address, Department of Physics, Howard University, Washington, DC 20059}
} \\
\end{center}
\bigskip\bigskip 
\begin{abstract} 
In a previous paper we extended the Lorentz group to include a set of
\textit{Dirac boosts} that give a direct correspondence with a set of
generators which for spin 1/2 systems are proportional to the Dirac matrices.  The
group is particularly useful for developing general linear wave equations
beyond spin 1/2 systems.  In this paper we develop explicit group properties
of this extended Lorentz group to obtain group parameters that will be
useful for physical calculations for systems which might manifest the
group properties.  This group is a subgroup of an extended Poincare group,
whose structure will be developed in a subsequent paper.

\end{abstract} 
\end{titlepage} 
\starttext \baselineskip=17.63pt \setcounter{footnote}{0} 
\setcounter{equation}{0} 
\section{ Introduction} 
\tab 
When developing general, non-perturbative models for physical systems
that satisfy the expected general cluster
decomposition properties for a multiparticle system,
one finds it particularly useful to have fields whose energy-momentum
equations satisfy linear dispersions in the manner of the Dirac equation\cite{Dirac}.
In previous papers\cite{jlspinor}\cite{xPoincare} we have developed a
set of group generators that satisfy this criterion.  In particular, we demonstrated
the Dirac matrices as proportional to the $\Lambda={1 \over 2}$ representation
of the group, as well as explicitly demonstrated the $\Lambda=1$ representation
of the group.

In this paper, we will develop the group structure elements for the extended Lorentz
group developed in reference\cite{jlspinor}.  In a subsequent paper, we will
develop the group structure elements for the extended Poincare group developed
in reference\cite{xPoincare}.  The particular representation being developed
as the covering group will be seen to be a subgroup of SL(4,C). 
We will explicitly calculate those group structure
elements that will be relavant for calculations involving systems which have
the extended Lorentz group as a local gauge symmetry.

\setcounter{equation}{0} 
\section{Group Theoretic Conventions} 
\tab 
In order to establish our conventions, we will briefly define the
group parameters that will be calculated in this paper.  In terms
of the group elements $\mathcal{M}$, a vector representation $S$
satisfies
\be
S(\mathcal{M}_2) \, S(\mathcal{M}_1) \: = \:
S(\Phi(\mathcal{M}_2;\mathcal{M}_1))
\label{representation}
\ee
where $\Phi(\mathcal{M}_2;\mathcal{M}_1)$ is the group composition rule. 
The generators of infinitesimal transformations are given by
\be
\left . i \, \mathbf{X}_r \: \equiv \: 
{\partial \over \partial \mathcal{M}^r} \, S(\mathcal{M}) \right |_{\mathcal{M}=\mathcal{I}} .
\label{generators}
\ee
Lie structure matrices can be defined by
\be
\mathbf{\Theta}_r ^s (\mathcal{M}) \: \equiv  \: 
\left . {\partial \Phi^s (\mathcal{M'};\mathcal{M})  \over 
\partial \mathcal{M'}^r } \right |_{\mathcal{M'}=\mathcal{I}}
\label{LieStructure}
\ee
from which one can obtain the group structure constants
\be
c_{sn}^m \: \equiv  \: 
\left . {\partial \over \partial \mathcal{M}^n} \mathbf{\Theta}_s ^m (\mathcal{M})
 \right |_{\mathcal{M}=\mathcal{I}} -
\left . {\partial \over \partial \mathcal{M}^s} \mathbf{\Theta}_n ^m (\mathcal{M})
 \right |_{\mathcal{M}=\mathcal{I}}
\ee
for the Lie algebra
\be
\left [ \mathbf{X}_r , \mathbf{X}_s  \right ] \: = \: -i c_{rs} ^m \mathbf{X}_m .
\ee
The generators transform under the representations of the group
as given by the relation
\be
S(\mathcal{M}^{-1}) \, \mathbf{X}_r \, S(\mathcal{M}) \: = \:
{\mathbf{O}\llap+}_r ^s (\mathcal{M}) \: \mathbf{X}_s
\label{Opluscalculate}
\ee
where the matrices \Oplus given by
\be
{\mathbf{O}\llap+}_r ^s (\mathcal{M}) \: \equiv \:
\left . {\partial \over \partial \mathcal{M'}^r}
\Phi^s (\mathcal{M}^{-1} \, ; \, \Phi(\mathcal{M'} \, ; \, \mathcal{M}))
\right | _{\mathcal{M'}=\mathcal{I}}
\ee
form a fundamental representation of the group. 

We will ultimately want to examine the behavior of systems
which have a local gauge symmetry under the group being
examined.  This means that we should examine derivatives
of the representations.  The relation
\be
\mathbf{\Theta}_r ^s (\mathcal{M}) {\partial \over \partial \mathcal{M}^s} 
S(\mathcal{M}) \: = \: i \, \mathbf{X}_r  \, S(\mathcal{M})
\ee
follows directly from Equations \ref{representation} and \ref{generators}
This means that if we want to construct a gauge covariant derivative 
$\mathbf{D}_\mu \: = \: \mathbf{1} \partial_\mu - i A_\mu ^r \, \mathbf{X}_r $
under the transformation $S$ using a gauge field $A_\mu ^r$,
then the gauge field will have a component $a_\mu ^r$ satisfying\cite{Mickens}\cite{Morrison}
\be
\begin{array}{l}
\partial_\mu \mathcal{M}^s \: = \: a_\mu ^r (\mathcal{M}) \,
\mathbf{\Theta}_r ^s (\mathcal{M}) \\ \\
A_\mu ^r (\vec{x}:\mathcal{M}) \: = \: a_\mu ^r (\mathcal{M}) +
A_\mu ^r (\vec{x}:\mathcal{I}) .
\end{array}
\ee
Under an infinitesimal gauge transformation, the gauge field then transforms
in the usual manner
\be
\delta A_\mu ^r \: \cong \: \partial _\mu (\delta \mathcal{M}^r) \, - \, 
\delta \mathcal{M}^s \, c_{s m} ^r \, A_\mu ^m .
\ee

We will therefore calculate the group matrix elements \Oplus  and  $\mathbf{\Theta}$ 
that determine the transformation properties of generators and gauge
fields for this group.  In subsequent papers, these elements will be used
in calculations involving dynamical models for physical systems.

\setcounter{equation}{0} 
\section{Lorentz Group Structure} 
\tab 
The Lorentz group is a subgroup of the extended Lorentz group being developed. 
We will first construct the desired elements for the Lorentz group.

\subsection{Pure Rotation Subgroup} 
\tab 
The rotation subgroup representation is given by
\be
\mathbf{R}(\underline{\theta}_2) \mathbf{R}(\underline{\theta}_1) \: = \:
\mathbf{R}(\underline{\theta}_{(R)} (\underline{\theta}_2;\underline{\theta}_1))
\ee
in terms of the group composition element
$\underline{\theta}_{(R)} (\underline{\theta}_2;\underline{\theta}_1)$. 
The inverse element  is given by $\{ \underline{\theta} \}^{-1} = \{ -\underline{\theta} \}$.

We can determine the group structure of the fundamental representation in
a straightforward manner.  For SU(2) we can explicitly write the representation in the form
\be
\mathbf{R}(\underline{\theta}) \: = \: e^{i \underline{\theta} \cdot \underline{\sigma}/2} =
\mathbf{1} \, cos({\theta \over 2}) + i \hat{\theta} \cdot \underline{\sigma} \, sin({\theta \over 2})
\label{SU2}
\ee
The SU(2) composition rule satisfies
\be
\begin{array}{l}
cos({\theta_{(R)} (\underline{\theta}_2 ; \underline{\theta}_1) \over 2}) \: = \: 
cos({\theta_2 \over 2}) cos({\theta_1 \over 2}) -
\hat{\theta}_2 \cdot \hat{\theta}_1 sin({\theta_2 \over 2}) sin({\theta_1 \over 2}) \\ \\
\hat{\theta}_{(R)} sin({\theta_{(R)}  (\underline{\theta}_2 ; \underline{\theta}_1) \over 2}) \: = \:
\hat{\theta}_2  sin({\theta_2 \over 2}) cos({\theta_1 \over 2}) +
\hat{\theta}_1 cos({\theta_2 \over 2}) sin({\theta_1 \over 2}) +
\hat{\theta}_2 \times \hat{\theta}_1 sin({\theta_2 \over 2}) sin({\theta_1 \over 2}) 
\end{array}
\label{SU2composition}
\ee
By direct substitution, Equation \ref{SU2} gives the fundamental representation:
\be
\begin{array}{l}
{\mathbf{O}\llap+}_{J_k}^{J_m} (\underline{0},\underline{\theta}) \: = \: 
cos(\theta) \delta_{k,m}  + (1-cos(\theta)) \hat{\theta}_k \hat{\theta}_m +
sin(\theta) \hat{\theta}_j \epsilon_{jkm} \\ \\
{\mathbf{O}\llap+}_{K_k}^{K_m} (\underline{0},\underline{\theta}) \: = \: 
cos(\theta) \delta_{k,m}  + (1-cos(\theta)) \hat{\theta}_k \hat{\theta}_m +
sin(\theta) \hat{\theta}_j \epsilon_{jkm}
\end{array}
\ee

The Lie structure matrices can be calculated using Equation \ref{SU2composition}
by examining infinitesimal $\underline{\theta}_2$;
\be
\mathbf{\Theta}_r ^{(R) s} (\underline{\theta}) \: = \: \delta_{r,s} +
{\theta_k \over 2} \epsilon_{krs} + 
\left({\theta \over 2}cot({\theta \over 2}) - 1 \right )
(\delta_{r,s} - \hat{\theta}_r \hat{\theta}_s)
\ee

\subsection{Lorentz Boosts} 
\tab 
In general, sequential pure Lorentz boosts can be written in terms of a
single pure Lorentz boost and a pure rotation:
\be
\mathbf{L}(\underline{u}_2) \, \mathbf{L}(\underline{u}_1) \: \equiv \:
\mathbf{L}(\underline{u}_{(L)}(\underline{u}_2 \, ; \, \underline{u}_1)) \,
\mathbf{R}(\underline{\theta}_{(L)}(\underline{u}_2 \, ; \, \underline{u}_1))
\ee
The composition rule for the covering group SL(2,C) representation can be
expressed
\be
\mathbf{L}_{\underline{\beta}} \: = \: e^{\underline{\beta} \cdot \underline{\sigma}/2} =
\mathbf{1} \, cosh({\beta \over 2}) +  \hat{\beta} \cdot \underline{\sigma} \, sinh({\beta \over 2}) .
\ee
Defining the four-velocity $\vec{u}$ using $u^0 = \sqrt{1+|\underline{u}|^2} \equiv 
cosh(\beta)$ and $\underline{u} \equiv \hat{\beta} sinh{\beta}$, the Lorentz boost takes the
form
\be
\mathbf{L}(\underline{u}) \: = \: 
 \sqrt{{u^0 + 1 \over 2}} \, \mathbf{1}  +   \sqrt{{u^0 - 1 \over 2}} \hat{u} \cdot \underline{\sigma}
\ee
The composition rule for pure boosts satisfies
\be
\begin{array}{l}
tan( { \theta_{(L)} ( \underline{u}_2 ; \underline{u}_1) \over 2 } ) \: = \:
{ | \underline{u}_2 \times \underline{u}_1 |    \over 
(u_2 ^0  +1) (u_1 ^0 +1) + \underline{u}_2 \cdot \underline{u}_1  } \\ \\
\hat{ \theta}_{(L)} ( \underline{u}_2 ; \underline{u}_1)  \: = \:
{ \underline{u}_2 \times \underline{u}_1 \over | \underline{u}_2 \times \underline{u}_1 | } \\ \\
u_{(L)} ^0 ( \underline{u}_2 ; \underline{u}_1)  \: = \:
u_2 ^0 \, u_1 ^0 \, + \, \underline{u}_2 \cdot \underline{u}_1 \\ \\
\left ( \hat{u}_{(L)} ( \underline{u}_2 ; \underline{u}_1)  \, cos({\theta_{(L)} \over 2}) -
 \hat{u}_{(L)} ( \underline{u}_2 ; \underline{u}_1) \times \hat{\theta}_{(L)} \, 
sin({\theta_{(L)} \over 2})    \right )  \sqrt{{u_{(L)}^0 -1 \over 2}} \: = \: \\
\quad \quad \quad \quad \quad \quad 
\hat{u}_2 \sqrt{{u_2 ^0 -1 \over 2}} \sqrt{{u_1 ^0 +1 \over 2}} + 
\hat{u}_1 \sqrt{{u_2 ^0 +1 \over 2}} \sqrt{{u_1 ^0 - 1 \over 2}}
\end{array}
\ee

\subsection{General Lorentz Transformations} 
\tab 
Lorentz boosts and rotations form a group of transformations
whose representation will be given by the convention
$\mathcal{L}(\underline{u},\underline{\theta}) \equiv 
\mathbf{L}(\underline{u}) \mathbf{R}(\underline{\theta}) $. 
The group composition behavior is given by
\be
\mathcal{L}(\underline{u}_2,\underline{\theta}_2) \, 
\mathcal{L}(\underline{u}_1,\underline{\theta}_1) \: = \:
\mathbf{L} \left ( \underline{u}_{(L)} (\underline{u}_2; 
R ( \underline{\theta}_2) \underline{u}_1) \right ) \,
\mathbf{R} \left ( \underline{\theta}_{(R)}(
\underline{\theta}_{(L)} (\underline{u}_2; R(\underline{\theta}_2) \underline{u}_1);
\underline{\theta}_{(R)}(\underline{\theta}_2 ; \underline{\theta}_1)) \right ) .
\ee
The inverse element of this representation of the Lorentz group satisfies
\be
\{ \underline{u}, \underline{\theta} \}^{-1} \: = \:
 \{ -R(-\underline{\theta})\underline{u}, -\underline{\theta}   \}
\ee

The fundamental representation matrix elements can be constructed
using Equation \ref{Opluscalculate} to obtain
\be
\begin{array}{l}
{\mathbf{O}\llap+}_{J_k}^{J_m} (\underline{u},\underline{0}) \: = \: 
u^0 \delta_{k,m}  + (1-u^0) \hat{u}_k \hat{u}_m \\ \\
{\mathbf{O}\llap+}_{J_k}^{K_m} (\underline{u},\underline{0}) \: = \: 
u_j \epsilon_{jkm} \\ \\
{\mathbf{O}\llap+}_{K_k}^{J_m} (\underline{u},\underline{0}) \: = \: 
-u_j \epsilon_{jkm} \\ \\
{\mathbf{O}\llap+}_{K_k}^{K_m} (\underline{u},\underline{0}) \: = \: 
u^0 \delta_{k,m}  + (1-u^0) \hat{u}_k \hat{u}_m 
\end{array}
\ee
The general matrix is then constructed from a rotation and sequential
boost using
\be
{\mathbf{O}\llap+}_r ^s (\underline{u},\underline{\theta}) \: = \: 
{\mathbf{O}\llap+}_r ^m (\underline{u},\underline{0}) \,
{\mathbf{O}\llap+}_m ^s (\underline{0},\underline{\theta}) \: = \:
{\mathbf{O}\llap+}_r ^{(L) \, m} (\underline{u}) \,
{\mathbf{O}\llap+}_m ^ {(R) \, s} (\underline{\theta}) .
\ee
In addition, the Lie structure matrices can be shown to be given by
\be
\begin{array}{l}
\mathbf{\Theta}_{\theta_k} ^{(\mathcal{L}) \theta_j} (\underline{u} ,  \underline{\theta}) \: = \:
\mathbf{\Theta}_k ^{(R) j} (  \underline{\theta}) \\ \\
\mathbf{\Theta}_{u_k} ^{ (\mathcal{L}) \theta_j} (\underline{u} ,  \underline{\theta}) \: = \:
\mathbf{\Theta}_{u_k} ^{(L) \theta_m} (  \underline{u} ) 
\mathbf{\Theta}_m ^{(R) j} (  \underline{\theta}) \\
\mathbf{\Theta}_{u_k} ^{(L) \theta_j} (  \underline{u}) \: = \:
-{u_m \over u^0 + 1} \, \epsilon_{mkj} \\ \\
\mathbf{\Theta}_{\theta_k} ^{(\mathcal{L}) u_j} (\underline{u} ,  \underline{\theta}) \: = \:
\mathbf{\Theta}_{\theta_k} ^{(L) u_j} (\underline{u} ) \, = \,
u_m \, \epsilon_{mkj} \\ \\
\mathbf{\Theta}_{u_k} ^{(\mathcal{L}) u_j} (\underline{u} ,  \underline{\theta}) \: = \:
\mathbf{\Theta}_{u_k} ^{(L) u_j} (\underline{u} ) \, = \,
u^0 \, \delta_{k,j}
\end{array}
\ee

To complete this section, we will establish our convention for the
Lorentz transformation matrices on four-vectors. 
Define $\mathcal{R}_\mu ^\nu$ and $\mathcal{L}_\mu ^\nu$
which act on (covariant) 4-vectors
according to $\Lambda_\mu ^\nu \omega_\nu =\omega_\mu '$
\be
\begin{array}{l}
\mathcal{R}_k ^m (\underline{\theta}) \: = \:
cos(\theta) \delta_{k,m}  + (1-cos(\theta)) \hat{\theta}_k \hat{\theta}_m +
sin(\theta) \hat{\theta}_j \epsilon_{jkm} \\ \\
\mathcal{R}_0 ^0 (\underline{\theta}) \: = \: 1 \\ \\ \\
\mathcal{L}_k ^m (\underline{u}) \: = \:
\delta_{k,m}  - (1-u^0) \hat{u}_k \hat{u}_m \\ \\
\mathcal{L}_0 ^m (\underline{u}) \: = \: -u_m \: = \: \mathcal{L}_m ^0 (\underline{u}) \\ \\
\mathcal{L}_0 ^0 (\underline{u} ) \: = \: u^0
\end{array}
\ee
The form of the infinitesimal 4-Lorentz generators
\be
\begin{array}{c}
\left ( \mathcal{J}_m \right ) _\mu ^\nu \, \equiv \,
\left . {\partial \over \partial \theta_m} \mathcal{R} _\mu ^\nu (\underline{\theta})
\right |_{\underline{\theta}=\underline{0}} \\ \\
\left ( \mathcal{K}_m \right ) _\mu ^\nu \, \equiv \,
\left . {\partial \over \partial u_m} \mathcal{L} _\mu ^\nu (\underline{u})
\right |_{\underline{u}=\underline{0}}
\end{array}
\ee
has non-vanishing elements given by
\be
\begin{array}{c}
\left ( \mathcal{J}_m \right ) _j ^k  \: = \: \epsilon_{mjk} \\ \\
\left ( \mathcal{K}_m \right ) _0 ^k \: = \: 
- \delta_{m,k} \: = \: \left ( \mathcal{K}_m \right ) _k ^0 .
\end{array}
\ee

\setcounter{equation}{0} 
\section{Extended Lorentz Group Structure} 
\tab 
The primary purpose of this paper is the development of the
group structure for the extended Lorentz group developed in
a previous paper\cite{jlspinor}.  The covering group will
be developed as a subgroup of SL(4,C).

\subsection{Form of $\Lambda={1 \over 2}$ Matrices}
\tab
The forms of the lowest dimensional matrices of the extended
Lorentz group corresponding to $\Lambda={1 \over 2}$ 
can be expressed in terms of the Pauli spin matrices as shown below:
\be
\begin{array}{l l}
\mathbf{\Gamma^0} \,=\, {1 \over 2} \left( \begin{array}{r r}
\mathbf{1} & \mathbf{0} \\ \mathbf{0} & -\mathbf{1} \end{array} \right)
\,\equiv \, {1 \over 2} \mathbf{\gamma^0} \quad \quad &
\mathbf{\underline{J} } \,=\, {1 \over 2} \left( \begin{array}{r r}
\mathbf{\underline{\sigma} } & \mathbf{0} \\ \mathbf{0} & \mathbf{\underline{\sigma} } 
\end{array} \right) \\ \\
\mathbf{\underline{\Gamma}} \,=\, {1 \over 2} \left( \begin{array}{r r}
\mathbf{0} & \mathbf{\underline{\sigma} } \\ 
-\mathbf{\underline{\sigma} } & \mathbf{0} \end{array} \right) 
\,\equiv \, {1 \over 2} \mathbf{\underline{\gamma}} \quad \quad&
\mathbf{\underline{K}} \,=\, -{i \over 2} \left( \begin{array}{r r}
\mathbf{0} & \mathbf{\underline{\sigma} } \\ 
\mathbf{\underline{\sigma} } & \mathbf{0} \end{array} \right)
\end{array}
\ee
The $\Gamma^\mu$ matrices can directly be seen to be proportional
to a representation of the Dirac matrices\cite{Dirac}\cite{BjDrell} . 

\subsection{Dirac Boosts} 
\tab 
Group transformations parameterized by elements $\vec{\omega}$
conjugate to the generators $\Gamma^\mu$ will be referred to as
pure Dirac boosts.  In general, sequential pure Dirac boosts can
be written in terms of a single pure Dirac boost, a pure Lorentz boost,
and a pure rotation:
\be
\mathbf{W}(\vec{\omega}_2) \, \mathbf{W}(\vec{\omega}_1) \: \equiv \:
\mathbf{W}(\vec{\omega}_{(D)}(\vec{\omega}_2 \, ; \, \vec{\omega}_1)) \,
\mathbf{L}(\underline{u}_{(D)}(\vec{\omega}_2 \, ; \, \vec{\omega}_1)) \,
\mathbf{R}(\underline{\theta}_{(D)}(\vec{\omega}_2 \, ; \, \vec{\omega}_1)) .
\ee
To develop the composition rules for the lowest dimensional representation,
the four-vector magnitude and direction will be expressed
\be
\begin{array}{l}
\omega \:\equiv \: \sqrt{-\vec{\omega} \cdot \vec{\omega}} \\ \\
\vec{\omega} \: = \: \omega \, \vec{q} .
\end{array}
\ee
This allows the Dirac representation to be expressed in terms of the
Dirac matrices in the form
\be
e^{i \vec{\omega} \cdot \vec{\gamma}/2} \: = \: \left \{
\begin{array}{l l}
\mathbf{1} cosh({\omega \over 2}) + i \vec{q} \cdot \vec{\gamma} sinh({\omega \over 2}) &
\vec{q} \cdot \vec{q} = +1 \\ \\
\mathbf{1} + i \vec{\omega} \cdot \vec{\gamma} /2 &
\vec{\omega} \cdot \vec{\omega} = 0 \\ \\
\mathbf{1} cos({\omega \over 2}) + i \vec{q} \cdot \vec{\gamma} sin({\omega \over 2}) &
\vec{q} \cdot \vec{q} = -1
\end{array}
\right .
\ee
For this representation, the composition elements satisfy
\be
\begin{array}{l}
cos({\omega_{(D)} \over 2}) \, \sqrt{{ u_{(D)} ^0 + 1\over 2}} \, cos({\theta_{(D)}\over 2}) \: = \:
cos({\omega_2 \over 2}) \, cos({\omega_1 \over 2}) \, + \,
\vec{q}_2 \cdot \vec{q}_1 \, sin({\omega_2 \over 2}) \, sin({\omega_1 \over 2}) , \\ \\
cos({\omega_{(D)} \over 2}) \, \sqrt{{ u_{(D)} ^0 + 1\over 2}} \, sin({\theta_{(D)}\over 2}) \,
\hat{\theta}_{(D)} \: = \: sin({\omega_2 \over 2}) \, sin({\omega_1 \over 2}) \,
\underline{q}_2 \times \underline{q}_1 , \\ \\
cos({\omega_{(D)} \over 2}) \, \sqrt{{ u_{(D)} ^0 - 1\over 2}} \, cos({\theta_{(D)}\over 2}) \,
\hat{u}_{(D)} + cos({\omega_{(D)} \over 2}) \, \sqrt{{ u_{(D)} ^0 - 1\over 2}} \, sin({\theta_{(D)}\over 2}) \,
\hat{\theta}_{(D)} \times \hat{u}_{(D)} \: = \: \\
\quad \quad \quad \quad \quad \quad \quad \quad \quad
sin({\omega_2 \over 2}) \, sin({\omega_1 \over 2}) \,
(\underline{q}_2 q_{1 \, 0} -\underline{q}_1 q_{2 \, 0}) , \\ \\
sin({\omega_{(D)} \over 2}) \, \sqrt{{ u_{(D)} ^0 + 1\over 2}} \, cos({\theta_{(D)}\over 2}) \, q_{(D) \,0} +
sin({\omega_{(D)} \over 2}) \, \sqrt{{ u_{(D)} ^0 - 1\over 2}} \, cos({\theta_{(D)}\over 2}) \, 
\underline{q}_{(D)} \cdot \hat{u}_{(D)} + \\
-sin({\omega_{(D)} \over 2})  \sqrt{{ u_{(D)} ^0 - 1\over 2}}  sin({\theta_{(D)}\over 2}) 
( \hat{\theta}_{(D)} \times \underline{q}_{(D)} ) \cdot \hat{u}_{(D)}  = 
q_{2 \, 0}  sin({\omega_2 \over 2})  cos({\omega_1 \over 2}) +
q_{1 \, 0}  cos({\omega_2 \over 2})  sin({\omega_1 \over 2}) , \\ \\
sin({\omega_{(D)} \over 2}) \, \sqrt{{ u_{(D)} ^0 + 1\over 2}} \, cos({\theta_{(D)}\over 2}) \, 
\underline{q}_{(D)} +
sin({\omega_{(D)} \over 2}) \, \sqrt{{ u_{(D)} ^0 - 1\over 2}} \, cos({\theta_{(D)}\over 2}) \, 
q_{(D) \, 0} \hat{u}_{(D)} + \\
sin({\omega_{(D)} \over 2})  \sqrt{{ u_{(D)} ^0 + 1\over 2}}  sin({\theta_{(D)}\over 2}) 
( \hat{\theta}_{(D)} \times \underline{q}_{(D)} ) +
sin({\omega_{(D)} \over 2})  \sqrt{{ u_{(D)} ^0 - 1\over 2}}  sin({\theta_{(D)}\over 2}) 
\, q_{(D) \, 0} \, \hat{\theta}_{(D)} \times\hat{u}_{(D)} \: = \\
\quad \quad \quad \quad \quad \quad \quad \quad
\underline{q}_{2}  sin({\omega_2 \over 2})  cos({\omega_1 \over 2}) +
\underline{q}_{1}  cos({\omega_2 \over 2})  sin({\omega_1 \over 2})
\end{array}
\ee
with constraints given by
\be
\begin{array}{l}
\hat{u}_{(D)} \cdot \hat{\theta}_{(D)} \: = \: 0 \\ \\
\underline{q}_{(D)} \cdot \hat{\theta}_{(D)} \: = \: 0 \\ \\
\sqrt{{ u_{(D)} ^0 - 1\over 2}} \, cos({\theta_{(D)}\over 2}) \, 
\underline{q}_{(D)} \times \hat{u}_{(D)} + \\
\sqrt{{ u_{(D)} ^0 - 1\over 2}} \, sin({\theta_{(D)}\over 2}) \, 
\underline{q}_{(D)} \cdot \hat{u}_{(D)} \, \hat{\theta}_{(D)} +
\sqrt{{ u_{(D)} ^0 + 1\over 2}} \, sin({\theta_{(D)}\over 2}) \, 
q_{(D) \, 0}  \: \hat{\theta}_{(D)} \: = \: 0
\end{array}
\ee

As was done for the Lorentz group,
the fundamental representation matrix elements can be constructed
using Equation \ref{Opluscalculate} to obtain
\be
\begin{array}{l}
{\mathbf{O}\llap+}_{J_k}^{J_s} (\vec{\omega},\underline{0},\underline{0}) \: = \: 
\left ( 1 + |\underline{q}|^2 (1-cos(\omega)) \right ) \delta_{k,s} - 
q_k q_s (1-cos(\omega) ) \\ \\
{\mathbf{O}\llap+}_{J_k}^{K_s} (\vec{\omega},\underline{0},\underline{0}) \: = \: 
 q_0 q_j (1-cos(\omega) ) \epsilon_{jks} \\ \\
{\mathbf{O}\llap+}_{J_k}^{\Gamma^0} (\vec{\omega},\underline{0},\underline{0}) \: = \: 0 \\ \\
{\mathbf{O}\llap+}_{J_k}^{\Gamma^s} (\vec{\omega},\underline{0},\underline{0}) \: = \: 
q_j sin(\omega)  \epsilon_{jks} \\ \\
{\mathbf{O}\llap+}_{K_k}^{J_s} (\vec{\omega},\underline{0},\underline{0}) \: = \: 
 q_0 q_j (1-cos(\omega) ) \epsilon_{jks} \\ \\
{\mathbf{O}\llap+}_{K_k}^{K_s} (\vec{\omega},\underline{0},\underline{0}) \: = \: 
\left ( cos(\omega) - |\underline{q}|^2 (1-cos(\omega)) \right ) \delta_{k,s} + 
q_k q_s (1-cos(\omega) ) \\ \\
{\mathbf{O}\llap+}_{K_k}^{\Gamma^0} (\vec{\omega},\underline{0},\underline{0}) \: = \: 
-q_k sin(\omega)  \\ \\
{\mathbf{O}\llap+}_{K_k}^{\Gamma^s} (\vec{\omega},\underline{0},\underline{0}) \: = \: 
-q_0 sin(\omega) \delta_{k,s} \\ \\
{\mathbf{O}\llap+}_{\Gamma^0}^{J^s} (\vec{\omega},\underline{0},\underline{0}) \: = \: 0  \\ \\
{\mathbf{O}\llap+}_{\Gamma^0}^{K^s} (\vec{\omega},\underline{0},\underline{0}) \: = \: 
-q_s sin(\omega)  \\ \\
{\mathbf{O}\llap+}_{\Gamma^0}^{\Gamma^0} (\vec{\omega},\underline{0},\underline{0}) \: = \: 
\left ( 1 + |\underline{q}|^2 (1-cos(\omega)) \right )  \\ \\
{\mathbf{O}\llap+}_{\Gamma^0}^{\Gamma^s} (\vec{\omega},\underline{0},\underline{0}) \: = \: 
q_0 q_s (1-cos(\omega))  \\ \\
{\mathbf{O}\llap+}_{\Gamma^k}^{J^s} (\vec{\omega},\underline{0},\underline{0}) \: = \: 
-q_j sin(\omega)  \epsilon_{jks} \\ \\
{\mathbf{O}\llap+}_{\Gamma^k}^{K^s} (\vec{\omega},\underline{0},\underline{0}) \: = \: 
q_0 sin(\omega) \delta_{k,s} \\ \\
{\mathbf{O}\llap+}_{\Gamma^k}^{\Gamma^0} (\vec{\omega},\underline{0},\underline{0}) \: = \: 
-q_0 q_s (1-cos(\omega))  \\ \\
{\mathbf{O}\llap+}_{\Gamma^k}^{\Gamma^s} (\vec{\omega},\underline{0},\underline{0}) \: = \: 
cos(\omega))  \delta_{k,s} - q_k q_s (1-cos(\omega) )  
\end{array}
\ee
The non-vanishing Lie structure matrices are given by
\be
\begin{array}{l}
\mathbf{\Theta}_{\omega_k} ^{(D) \theta_j} (\vec{\omega}) \: = \:
\epsilon_{jkm} \, q_m \, tan({\omega \over 2})\\ \\
\mathbf{\Theta}_{\omega_0} ^{(D) u_j} (\vec{\omega}) \: = \:
- q_j \, tan({\omega \over 2})\\ \\
\mathbf{\Theta}_{\omega_k} ^{(D) u_j} (\vec{\omega}) \: = \:
\delta_{j,k} \, q_0 \, tan({\omega \over 2})\\ \\
\mathbf{\Theta}_{\omega_\mu} ^{(D) \omega_\beta} (\vec{\omega}) \: = \:
{\omega \over 2} \, cot({\omega \over 2}) \, \delta _\beta ^\mu \, - \,
\left (1- {\omega \over 2} cot({\omega \over 2}) \right ) \, q_\beta \, q^\mu 
\end{array}
\ee

\subsection{Lorentz Transformations on Dirac Boosts}
\tab
We next examine the effect of Lorentz transformations on the Dirac boosts. 
Utilizing the Equation \ref{Opluscalculate} for this representation, we can read
off various elements of the fundamental representation.  For pure rotations
\be
\begin{array}{l}
{\mathbf{O}\llap+}_{J_k}^{J_m} (\vec{0},\underline{0},\underline{\theta}) \: = \: 
cos(\theta) \delta_{k,m}  + (1-cos(\theta)) \hat{\theta}_k \hat{\theta}_m +
sin(\theta) \hat{\theta}_j \epsilon_{jkm} \\ \\
{\mathbf{O}\llap+}_{K_k}^{K_m} (\vec{0},\underline{0},\underline{\theta}) \: = \: 
cos(\theta) \delta_{k,m}  + (1-cos(\theta)) \hat{\theta}_k \hat{\theta}_m +
sin(\theta) \hat{\theta}_j \epsilon_{jkm} \\ \\
{\mathbf{O}\llap+}_{\Gamma^\mu}^{\Gamma^\nu} (\vec{0},\underline{0},\underline{\theta}) \: = \: 
\mathcal{R}_\mu ^\nu (\underline{\theta})
\end{array}
\ee
and for pure Lorentz boosts
\be
\begin{array}{l}
{\mathbf{O}\llap+}_{J_k}^{J_m} (\vec{0},\underline{u},\underline{0}) \: = \: 
u^0 \delta_{k,m}  + (1-u^0) \hat{u}_k \hat{u}_m \\ \\
{\mathbf{O}\llap+}_{J_k}^{K_m} (\vec{0},\underline{u},\underline{0}) \: = \: 
-u_j \epsilon_{kjm} \\ \\
{\mathbf{O}\llap+}_{K_k}^{J_m} (\vec{0},\underline{u},\underline{0}) \: = \: 
u_j \epsilon_{kjm} \\ \\
{\mathbf{O}\llap+}_{K_k}^{K_m} (\vec{0},\underline{u},\underline{0}) \: = \: 
u^0 \delta_{k,m}  + (1-u^0) \hat{u}_k \hat{u}_m \\ \\
{\mathbf{O}\llap+}_{\Gamma^\mu}^{\Gamma^\nu} (\vec{0},\underline{u},\underline{0}) \: = \: 
\mathcal{L}_\mu ^\nu (\underline{u})
\end{array}.
\ee
The general fundamental transformation matrix is then given by
\be
{\mathbf{O}\llap+}_r ^s (\vec{\omega},\underline{u},\underline{\theta}) \: = \: 
{\mathbf{O}\llap+}_r ^n (\vec{\omega},\underline{0},\underline{0}) \,
{\mathbf{O}\llap+}_n ^m (\vec{0},\underline{u},\underline{0}) \,
{\mathbf{O}\llap+}_m ^s (\vec{0},\underline{0},\underline{\theta}) .
\ee

\subsection{Extended Group Transformations} 
\tab 
To complete our description of the group structure of this extended
Lorentz group, we will explicitly demonstrate the group composition
rule and Lie structure elements of the complete group.  The representation
we develop will be constructed by sequential pure rotation, Lorentz boost,
and Dirac boost: 
\be
\mathbf{M} (\vec{\omega}, \underline{u}, \underline{\theta}) \: \equiv \:
\mathbf{W}(\vec{\omega}) \, \mathbf{L}(\underline{u}) \, \mathbf{R} (\underline{\theta})
\ee
The group composition rule then defines elements in this same manner
\be
\begin{array}{l}
\mathbf{M} (\vec{\omega}_2, \underline{u}_2, \underline{\theta}_2) \,
\mathbf{M} (\vec{\omega}_1, \underline{u}_1, \underline{\theta}_1) \: \equiv \: \\
\mathbf{W}(\vec{\omega}(\vec{\omega}_2, \underline{u}_2, \underline{\theta}_2 \, ; \,
\vec{\omega}_1, \underline{u}_1, \underline{\theta}_1) ) \, 
\mathbf{L}(\underline{u}(\vec{\omega}_2, \underline{u}_2, \underline{\theta}_2 \, ; \,
\vec{\omega}_1, \underline{u}_1, \underline{\theta}_1)) \, 
\mathbf{R} (\underline{\theta}(\vec{\omega}_2, \underline{u}_2, \underline{\theta}_2 \, ; \,
\vec{\omega}_1, \underline{u}_1, \underline{\theta}_1))
\end{array}
\ee
The inverse element can be shown to be given by
\be
\{\vec{\omega}, \underline{u}, \underline{\theta} \}^{-1} \: = \:
 \{ -\mathcal{R}(-\underline{\theta}) \, \mathcal{L}(-\underline{u}) \, \vec{\omega} , 
-R(-\underline{\theta})\underline{u}, -\underline{\theta}   \}
\ee
where $(\Lambda \vec{\omega})_\mu=\Lambda_\mu ^\nu \omega_\nu$. 
The group composition elements can be expressed using previously
constructed functions in terms of pure boosts and rotations:
\be
\begin{array}{l}
\vec{\omega}(\vec{\omega}_2, \underline{u}_2, \underline{\theta}_2 \, ; \,
\vec{\omega}_1, \underline{u}_1, \underline{\theta}_1) \: = \: 
\vec{\omega}_{(D)} \left ( 
\vec{\omega}_2 \, ; \, \mathcal{L}(\underline{u}_2) \mathcal{R}(\underline{\theta}_2) 
\vec{\omega}_1   \right ) \\ \\
\underline{u}(\vec{\omega}_2, \underline{u}_2, \underline{\theta}_2 \, ; \,
\vec{\omega}_1, \underline{u}_1, \underline{\theta}_1) \: = \:  \\
\underline{u}_{(L)} \left ( 
\underline{u}_{(D)} \left ( 
\vec{\omega}_2 \, ; \, \mathcal{L}(\underline{u}_2) \mathcal{R}(\underline{\theta}_2) 
\vec{\omega}_1   \right ) \, ; \,
\mathcal{R} (
\underline{\theta}_{(D)} \left ( 
\vec{\omega}_2 \, ; \, \mathcal{L}(\underline{u}_2) \mathcal{R}(\underline{\theta}_2) 
\vec{\omega}_1   \right )  ) \,
\underline{u}_{(L)} \left ( \underline{u}_2 \, ; \,  \mathcal{R}(\underline{\theta}_2) 
\underline{u}_1  \right )  \right ) \\ \\
\underline{\theta}(\vec{\omega}_2, \underline{u}_2, \underline{\theta}_2 \, ; \,
\vec{\omega}_1, \underline{u}_1, \underline{\theta}_1) \: = \:  \\
\underline{\theta}_{(R)} \left ( 
\underline{\theta}_{(D)} \left ( 
\vec{\omega}_2 \, ; \, \mathcal{L}(\underline{u}_2) \mathcal{R}(\underline{\theta}_2) 
\vec{\omega}_1   \right ) \, ; \,
\underline{\theta}_{(R)} \left (
\underline{\theta}_{(L)} \left ( \underline{u}_2 \, ; \,  \mathcal{R}(\underline{\theta}_2) 
\underline{u}_1 \right ) \, ; \,
\underline{\theta}_{(R)} (\underline{\theta}_2 \, ; \, \underline{\theta}_1)
\right ) \right )
\end{array}
\label{ELGcomposition}
\ee

\subsection{Lie Structure Elements} 
\tab 
Finally, we can use the composition rules given in Equation \ref{ELGcomposition},
along with the definition given in Equation \ref{LieStructure}, to explicitly develop
the Lie structure matrices
\be
\begin{array}{l}
\mathbf{\Theta}_{\omega_\nu} ^{\omega_\mu} (\vec{\omega},
\underline{u}, \underline{\theta}) \: = \:
\mathbf{\Theta}_{\omega_\nu} ^{(D) \, \omega_\mu} (\vec{\omega}) \\ \\
\mathbf{\Theta}_{u_j} ^{\omega_\mu} (\vec{\omega},
\underline{u}, \underline{\theta}) \: = \:
\left ( \mathcal{K}_j  \right )_\mu ^\beta  \omega_\beta \\ \\
\mathbf{\Theta}_{\theta_j} ^{\omega_\mu} (\vec{\omega},
\underline{u}, \underline{\theta}) \: = \:
\left ( \mathcal{J}_j  \right )_\mu ^\beta  \omega_\beta \\ \\
\mathbf{\Theta}_{\omega_\nu} ^{u_k} (\vec{\omega},
\underline{u}, \underline{\theta}) \: = \:
\mathbf{\Theta}_{\omega_\nu} ^{(D) \, u_m} (\vec{\omega}) \,
\mathbf{\Theta}_{u_m} ^{(L) \, u_k} (\underline{u}) \, + \,
\mathbf{\Theta}_{\omega_\nu} ^{(D) \, \theta_m} (\vec{\omega}) \,
\left ( \mathcal{J}_m  \right ) _k ^s \, u_s \\ \\
\mathbf{\Theta}_{u_j} ^{u_k} (\vec{\omega},
\underline{u}, \underline{\theta}) \: = \:
\mathbf{\Theta}_{u_j} ^{(L) \, u_k} (\underline{u}) \\ \\
\mathbf{\Theta}_{\theta_j} ^{u_k} (\vec{\omega},
\underline{u}, \underline{\theta}) \: = \:
\left ( \mathcal{J}_j  \right ) _k ^s \, u_s \\ \\
\mathbf{\Theta}_{\omega_\nu} ^{\theta_k} (\vec{\omega},
\underline{u}, \underline{\theta}) \: = \:
\mathbf{\Theta}_{\omega_\nu} ^{(D) \, \theta_m} (\vec{\omega}) \,
\mathbf{\Theta}_{m} ^{(R) \, k} (\underline{\theta}) \\ \\
\mathbf{\Theta}_{u_j} ^{\theta_k} (\vec{\omega},
\underline{u}, \underline{\theta}) \: = \:
\mathbf{\Theta}_{u_j} ^{(L) \, \theta_m} (\underline{u}) \,
\mathbf{\Theta}_{m} ^{(R) \, k} (\underline{\theta}) \\ \\
\mathbf{\Theta}_{\theta_j} ^{\theta_k} (\vec{\omega},
\underline{u}, \underline{\theta}) \: = \:
\mathbf{\Theta}_{j} ^{(R) \, k} (\underline{\theta})
\end{array}
\ee
We will be able to utilize these matrices in general
gauge transformations for systems which have a local
gauge symmetry in this group.

\section{Conclusions}
\tab
We have demonstrated an explicit group representation of
the extended Lorentz group developed in reference\cite{jlspinor}. 
The fundamental representation matrices can be seen to indeed
form a group of transformations.  Lie structure matrices that are
needed for the homogeneous part of transformation of gauge
fields have also be explicitly calculated.  Further group properties,
such as vector recoupling coefficients to express combined systems
in terms of combinations of irreducible components, will be left for subsequent
study.

\section{Acknowledgements}
\tab
The author wishes to acknowledge the support of Elnora Herod and
Penelope Brown during the intermediate periods prior to and after
his Peace Corps service (1984-1988), during which time the bulk of this work was
accomplished.  In addition, the author wishes to recognize the
hospitality of the Department of Physics at the University of Dar
Es Salaam during the three years from 1985-1987 in which a substantial portion of
this work was done.

\end{document}